\documentclass[12pt]{article}\usepackage[hyperfootnotes=false]{hyperref}
\usepackage{epsfig}
\usepackage{float}
\usepackage{empheq}
\usepackage{bbold}

\usepackage{xspace}

\makeatletter

\@addtoreset{equation}{section}
\makeatother

\usepackage[utf8]{inputenc}
\usepackage{amsmath}
\usepackage{caption}
\usepackage{amsmath}
\usepackage{amssymb}
\usepackage{graphicx}

\newlength{\abstractwidth}
\renewcommand{\thefootnote}{\fnsymbol{footnote}}
\renewcommand{\thanks}[1]{\footnote{#1}} 
\newcommand{\starttext}{
\setcounter{footnote}{0}
\renewcommand{\thefootnote}{\arabic{footnote}}}

\newcommand{\be}{\begin{equation}}
\newcommand{\bea}{\begin{eqnarray}}
\newcommand{\eea}{\end{eqnarray}}
\newcommand{\beq}{\begin{equation}}
\newcommand{\ee}{\end{equation}}

	\newcommand*\widefbox[1]{\fbox{\hspace{2em}#1\hspace{2em}}}

	\def\dss{de Sitter space}
	\def\dsp.{de Sitter space.}
	\def\eq{&=&}

	\def\ra{\rangle}
	\def\simleq{\; \raise0.3ex\hbox{$<$\kern-0.75em
			\raise-1.1ex\hbox{$\sim$}}\; }
	\def\simgeq{\; \raise0.3ex\hbox{$>$\kern-0.75em
			\raise-1.1ex\hbox{$\sim$}}\; }
	
	\def\bi{\begin{itemize}}
		\def\ei{\end{itemize}}

	\def\CC{{\cal{C}}}

	\def\CP{{\cal{P}}}

	\def\Tr{\rm Tr \it}
	\def\bsub{ \begin{subequations}
			\begin{empheq}[box=\widefbox]{align}  }
			\def\esub{ \end{empheq}
	\end{subequations}}
	
	\def\1{\(  \mathbb{1} \)}

	\def\bn{\bigskip \noindent}

	\def\dk{${\rm DSSYK_{\infty}}$}
	
	\makeatletter
	\g@addto@macro\normalsize{%
		\setlength\abovedisplayskip{10pt}
		\setlength\belowdisplayskip{20pt}
		\setlength\abovedisplayshortskip{10pt}
		\setlength\belowdisplayshortskip{20pt}
	}
	\makeatother
	
	\usepackage{color}

	\usepackage{authblk}
	\title{\Large \bf Why do we Need Observers? \\ Spontaneous Breaking of  Time-Reversal  in de Sitter Space.}

	\author[1,2]{\Large Leonard Susskind}
	\affil[1]{LITP and Department of Physics, Stanford University, Stanford, CA 94305-4060, USA \vspace{1em}}
	\affil[2]{Google, Mountain View, CA, USA}
	\date{}
	\usepackage{upgreek}

\begin{document}
		
		\begin{titlepage}
			\maketitle
			
			\begin{abstract}
			\Large
			In this paper I explain the relation between the need for observers in de Sitter space and the spontaneous breakdown of time-reversal symmetry.

 \end{abstract}

		\end{titlepage}
		
		\rightline{}
		\bigskip
		\bigskip\bigskip\bigskip\bigskip
		\bigskip
		
		\starttext \baselineskip=17.63pt \setcounter{footnote}{0}

	\LARGE

		\tableofcontents
		
\section{The Need for Observers}

In this note I will address the need for observers in de Sitter space  \cite{Chandrasekaran:2022cip}, especially in  the holography of a static patch. Some authors have interpreted this need as saying that without an observer quantum mechanics makes no sense. I will take a different view:   Without an observer the semiclassical limit  makes no sense, but quantum mechanics is perfectly OK. The holographic framework describing a static patch with or without an observer is quantum mechanics but in order to reproduce 
  semiclassical correlations the usual maximally mixed density matrix \cite{Chandrasekaran:2022cip}  must be modified. The modification requires the projection onto states with an appropriate feature at the pode. That feature can be called an observer but more important, it should include a ``forward-going clock" \cite{Susskind:2023rxm}. 
  
  Observers are lumps of matter that obey the same laws as any other lump. They occur very rarely with a probability,
  \be 
  \CP_o = e^{-M_o/T_{gh}}
  \label{prob}
  \ee
where the subscript $o$ means observer and $T_{gh}$ is the Gibbons-Hawking temperature of de Sitter space. As improbable as they are, one can condition on the presence of an observer at the pode.

  \bn
  
 The maximally mixed density matrix, for example in \dk, is the unit matrix, appropriately normalized,  and expectation values are calculated as normalized traces. For finite entropy this prescription is perfectly consistent but does not reproduce semiclassical correlations in the large entropy  limit.

\be
\Tr  \ \phi(t_1) \phi(t_2) \neq \text{semiclassical correlator}  
\label{neq}
\ee

 The conjecture of \cite{Chandrasekaran:2022cip}
 can be stated in the following way: Let $\Pi$ be a projection operator onto a subspace that contains an observer at the pode. Replacing the maximally mixed density matrix by $\Pi$ will reproduce semiclassical correlations.

\bea 
\Tr \Pi  \phi(t_1) \phi(t_2) &=&    \Tr   \phi(t_1) \phi(t_2) \Pi \cr
\eq \Tr \Pi  \phi(t_1) \phi(t_2) \Pi  \cr 
&=& \text{semiclassical correlator}
\label{eq}
\eea

 In each case the correlations, including the factors of $\Pi,$ are calculated in a proper quantum mechanical theory such as \dk.

 \subsection{Spontaneous Symmetry Breaking }
 In this paper I will explain that the need for observers is due to a spontaneous symmetry breaking (SSB) and that the projection  $\Pi$ is a very familiar process that one always does when SSB  occurs. Begin by recalling some well known aspects of  SSB: As an example  take the case of the 2-D Ising model on a very large but finite $N\times N$ lattice. The Hamiltonian is,
 \be  
 H_I=\sum_{nn}Z_i Z_j +\epsilon \sum_i X_i
 \label{HI}
 \ee
 where the sum $\sum_{nn}$ is over nearest neighbors and $\epsilon$ is small.
 
For finite but arbitrarily large $N$ the true ground state is the $Z(2)$ invariant state which is a superposition of positive and negative globally magnetized states,
 \be  
 |\text{ground state}\rangle =  |++++\cdots\rangle +  |----\cdots\rangle
 \label{grs}
 \ee
Correlation functions in this state violate the basic field-theoretic  principle of cluster decomposition. The one-point functions $\langle Z_i \rangle$ are all zero but correlation functions $\langle Z_i Z_j\rangle$ are non-zero at arbitrarily large separation.

We generally do not encounter such states in nature, the reason being that they are unstable. We resolve the violation of clustering in a manner that can be described in several ways:
\begin{enumerate}
    \item A very distant spin, $Z_o,$ very far  from the region of interest is chosen to be the ``observer"  and all other spins are ``dressed" relative to that spin,
\be
\bar{Z}_i = Z_o Z_i
\label{drsspn}
\ee

The one point function $\langle \bar{Z}_i \rangle$ is non-zero and correlation functions $$\langle \bar{Z}_i \bar{Z}_j \rangle$$ satisfy cluster decomposition.
\item We introduce a perturbation $H_o$ that acts only on $Z_o.$
\be 
H = H_I +H_o  =\bar{Z}_i =  \sum_{nn}Z_i Z_j  +\epsilon X_i +hZ_o
\label{pert}
\ee
This has the effect of polarizing $Z_o$ but more importantly it polarizes all the spins. In other words the true ground state becomes either 
$$|++++++\cdots\rangle$$
or 
$$|------\cdots\rangle$$
depending on the sign of $h.$ In this state cluster decomposition is satisfied.
Note that the perturbation is in a sense very small in that it only acts on a single very distant spin. In the $N\to \infty$ limit that spin is infinitely far from the region of interest, i.e., finite $i.$ The non-clustering state \eqref{grs} is unstable with respect to very small perturbations.
\item
Introduce a projection operator which projects onto states in which 
$Z_o=+1$.Then in terms of the theory without an observer the clustered system can be defined by correlations of the form,
\bea
C(i,j) &=& \langle\text{ground state}|   \Pi \  Z_i Z_j  \ \Pi  |\text{ground state}\rangle \cr \cr
\eq \Tr   \Pi \  Z_i Z_j  \ \Pi
\label{prjgs}
\eea

\end{enumerate}

To summarize: There is nothing wrong with quantum mechanics without an observer.  
The problem is just that clustering is violated due to spontaneous symmetry breaking. To resolve the problem it is convenient to introduce the observer spin $Z_o$ and dress the observables relative to the observer.

\bn

One important fact is that the details of the observer-system are not important.
It could be a single spin or a collection of several spins. In the latter case $\Pi$ may project onto all the spins being $+$ or just a majority of them being $+.$ The correlations far from the observer are universal and independent of details.

\section{SSB in de Sitter Space}

By now there are a number of puzzles which are resolved by introducing an observer system at the pode of the static patch. I will focus on one simple example described in  \cite{Susskind:2023rxm}. Consider a two-point function of a Hermitian operator $A$ in a theory in which the density matrix is maximally mixed,
\be  
C(t) = \langle A(0) A(t)\ra  = \Tr A(0) A(t).
\label{AA=Tr}
\ee  
The imaginary part of $C(t)$ is zero. This follows from the fact that 
\be  
\text{Im } \ C(t) = \langle [A(0), A(t)]\ra
\label{Im=com}
\ee
and the trace of a commutator is zero.

On the other hand the imaginary part of a two-point correlator in \dss \ is generally not zero in the semiclassical limit. This is a puzzling apparent inconsistency between semiclassical gravity and the holographic principle in \dss.

  The resolution of this conflict was suggested in   \cite{Susskind:2023rxm} where, following  \cite{Chandrasekaran:2022cip},     I argued that semiclassical correlations require conditioning on the presence of an observer, or more precisely the presence of a forward-going-clock in order  to break the symmetry of time reversal. 

Let $ \Pi $ now represent a projection onto states that contain a lump of matter at the pode including  a forward-going-clock. While 
\be  
\text{Im } \Tr A(0) A(t) =0
\label{Im=0}
\ee 
the projected correlator is not real.
\be  
\text{Im }  \Tr \Pi  A(0) A(t)   \neq 0.  
\label{Imneq0} 
\ee

To relate this to SSB we note that $[A(0), A(t)]$ is odd under time reversal\footnote{The vanishing of the imaginary part of the two-point function is one of many discrepancies between correlations in the maximally mixed state and semiclassical correlations. As Witten has pointed out, n-point functions will be cyclic invariant in the maximally mixed state but not semiclassically. Projecting with $\Pi$ will generally break the cyclic invariance.}. As argued in  \cite{Susskind:2023rxm} we not only require  a clock to be present  
at the pode, but also we need to replace the holographic time $t$ by the clock time. We assume that the time indicated on the forward-going-clock $t_c$ is an operator in the full Holographic  Hilbert space  and  we replace $A(t) = e^{-iHt } A e^{iHt}$ by  \cite{Chandrasekaran:2022cip}

\be 
A(t_c) = e^{-iHt_c } A e^{iHt_c}.
\label{A(tc)}
\ee

In other words we dress $A$ to the physical forward-going clock. 

In the limit that the de Sitter entropy becomes infinite  (the flat-space limit)  the pode moves infinitely far from the horizon. Correlation functions at a finite distance from the horizon develop imaginary parts when dressed to the clock at the pode, despite the fact that  the clock is no more than a small number of particles, and more importantly it may be arbitrarily far from the horizon. This means that cluster decomposition is violated in the maximally mixed state.

To see that clustering fails for the maximally mixed state consider the operator 
$$\CC =[A(0), A(t)]$$ where $A$ is an operator near the horizon and $t$ is a small time interval. The expectation value of $\CC$ is zero in the maximally mixed state.

Next  consider 
\be  
\Tr \  \CC \Pi 
\label{CPi}
\ee
where $\Pi $  is a projector onto a state with an observer at the pode. This we expect to be non-zero. But since in the flat-space limit the pode is infinitely far from the horizon, non-vanishing of \eqref{CPi} constitutes a breakdown of clustering in the bulk theory.

\bn

The conjecture by now is obvious:

\bn

\it 
\bn 
Time reversal invariance in the de Sitter static patch is spontaneously broken in the maximally mixed state. A clock arbitrarily far from the horizon will break the symmetry and when dressed to that clock correlation functions  will agree with the semiclassical limit.
  \rm

\bn

There are two ways to think about the introduction of $\Pi$ into correlation functions. One way is to think of it as a change in the density matrix from maximally mixed to $\Pi$ itself.  But a more consistent way is to think of quantities like \eqref{Imneq0} as correlation functions involving $\Pi$ as well as $A,$ calculated in the holographic  theory with a maximally mixed  density matrix. In this way of thinking the basic framework of de Sitter holography including a maximal entangle state is kept intact and $\Pi$ is just another operator to include in correlation functions.

As in the Ising model the details of the observer are not important as long as it breaks time-reversal symmetry.

\section*{Acknowledgement}

I have benefited from many discussions with Ying Zhao.

	\end{document}